\documentclass{ws-p10x7mio}

\begin{document}

\title{Study of Hadronic Decays of the Z Boson at LEP}

\author{Alessandro de Angelis}

\address{
Dipartimento di Fisica dell'Universit\`a di Udine and INFN Trieste, Via delle Scienze 208, I-33100 Udine, Italy -- E-mail: deangelis@ud.infn.it}

\twocolumn[\maketitle\abstract{
This report summarizes four recent papers 
on the characteristics of the hadronic decays of the Z 
by the LEP collaborations 
ALEPH, DELPHI and OPAL.}]

\section{Introduction}
Each of the four LEP experiments has recorded around 4 million hadronic Z 
decays, mostly in the period between 1990 and 1995. The analysis
of the main physics topics on this subject has been essentially completed, 
but a few specific points are still under investigation, especially in the QCD sector.

This report summarizes four recent papers 
on the characteristics of the hadronic decays of the Z,
submitted to this conference 
by the LEP collaborations 
OPAL\cite{opal}, 
ALEPH\cite{alepho,alepha}
and DELPHI\cite{delphi}.

\section{Charged multiplicities in Z decays 
into u, d, and s quarks}

Flavour independence is a fundamental property of  
QCD: a breaking of the flavour symmetry should only occur due to 
(calculable) mass effects. 
An observable for testing flavour independence is the multiplicity of charged 
hadrons in jets originating from quarks of a specific flavour. 
For the light ($u$, $d$, and 
$s$) quarks the mass effects are expected to be 
negligible at LEP energies\cite{ale}. 

OPAL\cite{opal} presents a new, high statistics investigation of the 
flavour dependence 
of the strong interaction based on the mean charged multiplicity determined separately 
for events of primary $u$, $d$, $s$ quarks in hadronic Z decays. 
 
In order to identify the flavour of the primary quark, 
the leading particle 
effect is exploited. This effect is based on the correlation between the flavour of the primary 
quark and the type of the hadron carrying the largest momentum.
 Three different selections are based on K$^0_S$, K$^\pm$ 
and 
highly energetic charged particles. Leading K$^0_S$ and 
K$^\pm$ are likely to contain 
a primary $s$ quark from the Z decay. 
To a lesser extent, due to the requirement 
of $s\bar{s}$ quark pair creation during hadronization, leading K$^0_S$ and
K$^\pm$
 should also emerge 
from primary $d$ and $u$ quarks, respectively. 
In a sample with large momentum (non identified) charged particles,  
events of primary $u$, $d$ and $s$ quarks are expected to 
be found in approximately equal proportions. 

By requiring a K$^0_S$ with momentum fraction
$x_p = p/p_{beam}$ larger than 0.4, 19,359 events are
selected with 58\% Z$\rightarrow s\bar{s}$ and  16\% Z$\rightarrow d\bar{d}$.
By requiring a K$^\pm$ with $x_p > 0.5$, 18,979 events are
selected with 54\% Z$\rightarrow s\bar{s}$ and  22\% Z$\rightarrow u\bar{u}$.
By requiring a charged particle with $x_p > 0.7$, 27,909 events are
selected almost equally populated in 
$u\bar{u}$, $d\bar{d}$ and $s\bar{s}$, but with only 4\% heavy quarks. 

The presence of a large momentum particle biases the multiplicity. To reduce this bias,
the multiplicity measurement was performed in the hemisphere opposite to 
the one containing the particle used for the
tagging.
The final 
corrected multiplicities are 
\begin{eqnarray*}
<n>_{u\bar{u}} & = & 17.77 \pm 0.52 (stat) ^{+0.86}_{-1.20} (sys)\\ 
<n>_{d\bar{d}} & = & 21.44 \pm 0.69 (stat) ^{+1.46}_{-1.17} (sys)\\
<n>_{d\bar{s}} & = & 20.02 \pm 0.14 (stat) ^{+0.39}_{-0.37} (sys) \, ,
\end{eqnarray*}
where the $u\bar{u}$ and $d\bar{d}$ multiplicities are
statistically anti­correlated (about -89\%). The systematic error is dominated by uncertainties 
on the fragmentation.
The multiplicities are consistent 
in $1.8\sigma$ for the $ud$ case,  
in $1.5\sigma$ for $us$, and in $0.9\sigma$ for $ds$.
The world average of the charge multiplicity\cite{pdg} (for all flavours)
is $21.07\pm0.11$.

The multiplicities were transformed into $\alpha_S$ ratios using a 
NLLA calculation\cite{webber}.
Ratios of $\alpha_S$ 
for the light quark flavours 
were found consistent with 1 
at a precision of 5 to 9\%, superior to earlier
investigations.

\section{Results on identified hadrons}

 The description of the hadronization process in QCD 
is connected with  
confinement and requires nonperturbative 
methods which are not available. 
Measurements of identified hadron spectra in 
$e^+e^-$ 
improve the understanding of hadronization, and
allow tuning the free parameters in the
QCD inspired Monte Carlo models. 
Being closer to the main event,
vectors and higher order resonances are of particular interest.

More insights into the hadronization process may be obtained from the analysis 
of quark and gluon jets separately.

\subsection{Production of $\omega(782)$}

ALEPH\cite{alepho} presents
a measurement of the inclusive momentum distributions 
of the $\omega(782)$  meson. 
 The $\omega$ cross section is extracted using the invariant mass
distribution of  $\pi^+\pi^-\pi^0$ triplets. 

The average $\omega$ rate per event  
is 
\begin{displaymath} 
<\omega> = 0.996 \pm 0.032 (stat) \pm 0.056 (sys) \, ,
\end{displaymath}
below 
the predictions of JETSET\cite{lund} 7.4 (1.31) and HERWIG\cite{herwig} 6.1 
(1.17) over the full momentum spectrum. This 
measurement improves 
substantially the world average\cite{pdg} of $1.08\pm0.09$. 

The muonic branching fraction of the $\omega(782)$ is measured for the first time: $BR(\omega \rightarrow \mu^+\mu^-) = (9.8\pm2.9\pm1.1)\times10^{-5}$.

\subsection{Production of $\pi^0$, $\eta$, $\eta'(958)$, K$^0_S$ and $\Lambda$ 
in 2-­ and 3-­jet events}

For isoscalar mesons ($\eta$, $\eta'(958)$, $\omega(782)$, $\phi(1020$)), 
some theoretical models 
predict a special enhancement in gluon jets compared to quark jets. 
L3\cite{lepo} found that the measured momentum spectrum in
gluon jets is harder than in 
HERWIG and JETSET.

In a new analysis, ALEPH\cite{alepha} 
measures the production rates of 
 $\pi^0$, $\eta$, $\eta'(958)$, K$^0_S$ and $\Lambda$ 
in hadronic events, two-­jet events and each jet of three-­jet events.

Jets are clustered using  
Durham\cite{durham} 
with $y_{cut}$ = 0.01; this classifies 31\% of the events as 3­-jet.
The lowest energy jet originates from a gluon with 71\% probability. 

The $\pi^0$ and $\eta$ mesons are analyzed using the $\gamma\gamma$ 
decay channel, and
the $\eta'$ using
$\eta' \rightarrow \eta \pi^+\pi^-$.
K$^0_S$ 
are reconstructed from their decay into $\pi^+\pi^-$, 
and $\Lambda$s into $p\pi^-$.

The spectra for $\pi^0$ are reasonably reproduced by JETSET 
and HERWIG. $\eta$ and $\eta'(958)$ are well 
reproduced by JETSET for quark jets and gluon jets. Therefore, the 
JETSET description of gluon fragmentation into 
isoscalar mesons is in agreement with 
the experiment. HERWIG shows a too steep 
dependence on $x_p$ for $\eta$ spectra in two-­jet events and each jet of 
three-­jet events.

The spectra for K$^0_S$ and
$\Lambda$ are well reproduced by 
JETSET and ARIADNE\cite{ariadne}, while HERWIG fails.

\section{Rapidity­-rank structure of $p\bar{p}$}

The baryon sector is not well modeled by current QCD--inspired Monte Carlos;
only the string fragmentation model
JETSET can reproduce the octet and decuplet rates.

In JETSET
hadrons results from breaks in the string  
stretched
between the primary quarks. 
Baryons are formed when 
diquark­-antidiquark pairs are created, 
and couple to an adjacent  
(anti)quark. 
A mechanism to attenuate the strict rapidity ordering coming from
this mechanism is the so-called {\em popcorn}: a meson  ``pops up''
inside a diquark--antidiquark pair\cite{lund}.

To study the relative occurrence of popcorn, DELPHI\cite{delphi} proposes a novel method based on the rapidity--rank structure of $p\bar{p}$ pairs, where
rapidity is defined with respect to the thrust axis.

Proton identification is provided by 
Cherenkov angle measurement from the RICH and ionization energy loss
in the TPC. 
The number of events with one $p$ and one $\bar{p}$ in each
hemisphere 
is 27.6 thousand. 
The background to this event 
sample can be determined from the number of events that have two 
$p$'s or two $\bar{p}$'s 
in a given 
hemisphere. These events, 10.1 thousand, result mainly from 
misidentifications, but  
also from non­correlated baryon pairs, and they are subtracted from
the signal.

The charged particles in each event are ordered in rapidity.
Two types of rapidity­-rank configurations for $p\bar{p}$ 
pairs 
are considered: (a)
when the $p$ and $\bar{p}$ 
are adjacent 
in rapidity; (b) when the $p$ and $\bar{p}$  
have one or more 
mesons between them. 
The ratio  
$R = N(pM\bar{p})/(N(p\bar{p})+N(pM\bar{p})$ is calculated, where
$N(p\bar{p})$ and $N(pM\bar{p})$ 
represent the number of rapidity-­rank configurations of each 
type in the data sample, and are implicitly a function of 
the minimum distance  $\Delta y_{min}$
between a baryon in the pair and a meson. 
The probability that a given rapidity configuration will represent the actual 
rank order on the string is enhanced as $\Delta y_{min}$ is made larger.

\begin{figure}
\mbox{\epsfxsize6cm\epsfysize6cm\epsffile{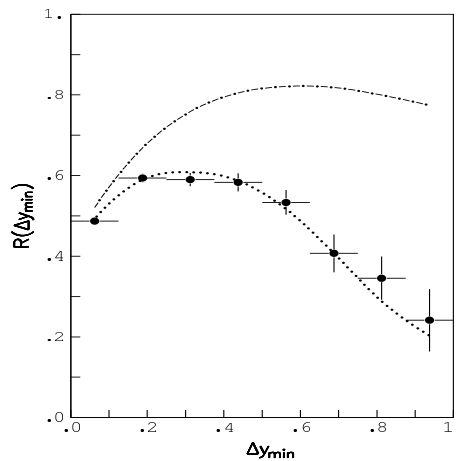}}
\caption{The relative amount, $R(\Delta y_{min})$, of the $pM\bar{p}$ 
configuration as a function of 
$\Delta y_{min}$, compared to the 
JETSET predictions in case of 
no popcorn (lower line) and full popcorn  
(upper line). Due to normalization, $\Delta y_{min}$ 
is multiplied by 2/3 for $p\bar{p}$.}
\label{lamsa2}
\end{figure}

The ratio $R(\Delta y_{min})$ for data is plotted in Figure \ref{lamsa2},
and compared to the predictions in the cases of no popcorn and 100\%
popcorn.
The data are consistent with no contribution 
from $p M \bar{p}$ configurations. 
An upper limit contribution of 15\% is 
determined at 90\% confidence level. 

Previous studies of the $\Lambda\bar{\Lambda}$ 
rapidity difference have 
claimed evidence for popcorn\cite{popcorn}. 
This might indicate the importance of dynamical effects 
not incorporated in JETSET or simply the inadequacy of the popcorn model, 
although no 
firm conclusion can be drawn yet.

\end{document}